\documentclass[aps,prl,twocolumn,superscriptaddress,showpacs,preprintnumbers,amsmath,amssymb]{revtex4}
\usepackage{graphicx} % Include figure files
\usepackage{dcolumn} % Align table columns on decimal point

\graphicspath{{ps}}

\begin{document}

%\vspace*{-3\baselineskip}
%\resizebox{!}{3cm}{\includegraphics{belle.eps}}

\preprint{\vbox{ \hbox{   }
                 %\vspace*{1.5cm}
                  \hbox{KEK Preprint 2007-70}
                  \hbox{BELLE Preprint 2008-1}
                 %\hbox{hep-ex nnnn, if available}
}}

\title{ \quad\\[0.5cm]  Study of $B \to \phi \phi K$ Decays}

%\input{author-conf2006}
%%%% >>>>> insert the authorlist here. BEFORE the abstract !!!!! <<<<<
%%%% >>>>> obtain the latest summer conference authorlist from the
%%%% >>>>> authorship confirmation web page

%\collaboration{Belle Collaboration}
%\noaffiliation
%%% Paper:    B-> phi phi K
%%% Journal:  Physical Review Letters
%%% Contacts: Y.-T. Shen (ytshen@hep1.phys.ntu.edu.tw)
%%%           P. Chang (pchang@phys.ntu.edu.tw)
%%% Non-responding authors or those who said NO are commented out.
%%% ====================================================================
%%% Click the RELOAD button on your web browser to see the updated file.
%%% ====================================================================
%%% Use \input{author} to insert this material into your latex file.
%%%%% Force institutions to appear in alphabetical order when typeset.
\affiliation{Budker Institute of Nuclear Physics, Novosibirsk}
\affiliation{Chiba University, Chiba}
\affiliation{University of Cincinnati, Cincinnati, Ohio 45221}
%%%\affiliation{Department of Physics, Fu Jen Catholic University, Taipei}
%%%\affiliation{Justus-Liebig-Universit\"at Gie\ss{}en, Gie\ss{}en}
\affiliation{The Graduate University for Advanced Studies, Hayama}
%%%\affiliation{Gyeongsang National University, Chinju}
\affiliation{Hanyang University, Seoul}
\affiliation{University of Hawaii, Honolulu, Hawaii 96822}
\affiliation{High Energy Accelerator Research Organization (KEK), Tsukuba}
%%%\affiliation{Hiroshima Institute of Technology, Hiroshima}
%%%\affiliation{University of Illinois at Urbana-Champaign, Urbana, Illinois 61801}
\affiliation{Institute of High Energy Physics, Chinese Academy of Sciences, Beijing}
\affiliation{Institute of High Energy Physics, Vienna}
\affiliation{Institute of High Energy Physics, Protvino}
\affiliation{Institute for Theoretical and Experimental Physics, Moscow}
\affiliation{J. Stefan Institute, Ljubljana}
\affiliation{Kanagawa University, Yokohama}
\affiliation{Korea University, Seoul}
\affiliation{Kyoto University, Kyoto}
\affiliation{Kyungpook National University, Taegu}
\affiliation{\'Ecole Polytechnique F\'ed\'erale de Lausanne (EPFL), Lausanne}
\affiliation{University of Ljubljana, Ljubljana}
\affiliation{University of Maribor, Maribor}
\affiliation{University of Melbourne, School of Physics, Victoria 3010}
\affiliation{Nagoya University, Nagoya}
\affiliation{Nara Women's University, Nara}
\affiliation{National Central University, Chung-li}
\affiliation{National United University, Miao Li}
\affiliation{Department of Physics, National Taiwan University, Taipei}
\affiliation{H. Niewodniczanski Institute of Nuclear Physics, Krakow}
\affiliation{Nippon Dental University, Niigata}
\affiliation{Niigata University, Niigata}
\affiliation{University of Nova Gorica, Nova Gorica}
\affiliation{Osaka City University, Osaka}
\affiliation{Osaka University, Osaka}
\affiliation{Panjab University, Chandigarh}
%%%\affiliation{Peking University, Beijing}
%%%\affiliation{University of Pittsburgh, Pittsburgh, Pennsylvania 15260}
%%%\affiliation{Princeton University, Princeton, New Jersey 08544}
%%%\affiliation{RIKEN BNL Research Center, Upton, New York 11973}
\affiliation{Saga University, Saga}
\affiliation{University of Science and Technology of China, Hefei}
\affiliation{Seoul National University, Seoul}
%%%\affiliation{Shinshu University, Nagano}
\affiliation{Sungkyunkwan University, Suwon}
\affiliation{University of Sydney, Sydney, New South Wales}
%%%\affiliation{Tata Institute of Fundamental Research, Mumbai}
\affiliation{Toho University, Funabashi}
\affiliation{Tohoku Gakuin University, Tagajo}
%%%\affiliation{Tohoku University, Sendai}
\affiliation{Department of Physics, University of Tokyo, Tokyo}
\affiliation{Tokyo Institute of Technology, Tokyo}
\affiliation{Tokyo Metropolitan University, Tokyo}
\affiliation{Tokyo University of Agriculture and Technology, Tokyo}
%%%\affiliation{Toyama National College of Maritime Technology, Toyama}
\affiliation{Virginia Polytechnic Institute and State University, Blacksburg, Virginia 24061}
\affiliation{Yonsei University, Seoul}
% \author{K.~Abe}\affiliation{High Energy Accelerator Research Organization (KEK), Tsukuba} % KEK
   \author{Y.-T.~Shen}\affiliation{Department of Physics, National Taiwan University, Taipei} % Taiwan
 \author{K.-F.~Chen}\affiliation{Department of Physics, National Taiwan University, Taipei} % Taiwan
 \author{P.~Chang}\affiliation{Department of Physics, National Taiwan University, Taipei} % Taiwan
   \author{I.~Adachi}\affiliation{High Energy Accelerator Research Organization (KEK), Tsukuba} % KEK
   \author{H.~Aihara}\affiliation{Department of Physics, University of Tokyo, Tokyo} % Tokyo
% \author{D.~Anipko}\affiliation{Budker Institute of Nuclear Physics, Novosibirsk} % BINP
   \author{K.~Arinstein}\affiliation{Budker Institute of Nuclear Physics, Novosibirsk} % BINP
% \author{T.~Aso}\affiliation{Toyama National College of Maritime Technology, Toyama} % Toyama
% \author{V.~Aulchenko}\affiliation{Budker Institute of Nuclear Physics, Novosibirsk} % BINP
   \author{T.~Aushev}\affiliation{\'Ecole Polytechnique F\'ed\'erale de Lausanne (EPFL), Lausanne}\affiliation{Institute for Theoretical and Experimental Physics, Moscow} % ITEP
% \author{T.~Aziz}\affiliation{Tata Institute of Fundamental Research, Mumbai} % Tata
% \author{S.~Bahinipati}\affiliation{University of Cincinnati, Cincinnati, Ohio 45221} % Cincinnati
   \author{A.~M.~Bakich}\affiliation{University of Sydney, Sydney, New South Wales} % Sydney
   \author{V.~Balagura}\affiliation{Institute for Theoretical and Experimental Physics, Moscow} % ITEP
% \author{Y.~Ban}\affiliation{Peking University, Beijing} % Peking
% \author{S.~Banerjee}\affiliation{Tata Institute of Fundamental Research, Mumbai} % Tata
% \author{E.~Barberio}\affiliation{University of Melbourne, School of Physics, Victoria 3010} % Melbourne
% \author{M.~Barbero}\affiliation{University of Hawaii, Honolulu, Hawaii 96822} % Hawaii
% \author{A.~Bay}\affiliation{\'Ecole Polytechnique F\'ed\'erale de Lausanne (EPFL), Lausanne} % Lausanne
% \author{I.~Bedny}\affiliation{Budker Institute of Nuclear Physics, Novosibirsk} % BINP
   \author{K.~Belous}\affiliation{Institute of High Energy Physics, Protvino} % Protvino
% \author{V.~Bhardwaj}\affiliation{Panjab University, Chandigarh} % Panjab
% ...
% Ljubljana
% \author{N.~Zwahlen}\affiliation{\'Ecole Polytechnique F\'ed\'erale de Lausanne (EPFL), Lausanne} % Lausanne
% \author{O.~Zyukova}\affiliation{Budker Institute of Nuclear Physics, Novosibirsk} % BINP
\collaboration{The Belle Collaboration}

\begin{abstract}
%%%
%%%
We report an observation of the decay $B^{\pm} \to \phi \phi K^{\pm}$ and  evidence for $B^{0} \to \phi \phi K^{0}$.
These results are based on a 414 fb$^{-1}$ data sample collected with the Belle detector at the KEKB asymmetric-energy $e^+e^-$ collider operating at the $\Upsilon(4S)$ resonance.
The branching fractions for these decay modes are measured to be $\mathcal{B}(B^{\pm} \to \phi \phi K^{\pm}) = (3.2^{+0.6}_{-0.5} \pm 0.3) \times 10^{-6}$ and $\mathcal{B}(B^{0} \to \phi \phi K^{0}) = (2.3^{+1.0}_{-0.7} \pm 0.2) \times 10^{-6}$ for $\phi \phi$ invariant mass below 2.85 GeV/$c^2$.
The corresponding partial rate asymmetry for the charged $B$ mode is measured
 to be 
$\mathcal{A}_{CP}(B^\pm\to \phi \phi K^\pm) = 0.01^{+ 0.19}_{-0.16}\pm 0.02$.
We also study the decays $B^\pm \to J/\psi K^\pm$ and $B^\pm \to \eta_c K^\pm$,
where the $J/\psi$ and $\eta_c$ decay to final states with 
four charged kaons. We find 
 $\mathcal{A_{CP}}(B^\pm \to \phi \phi K^\pm)$ with
the $\phi\phi$ candidates within the $\eta_c$ mass region is
$0.15^{+0.16}_{-0.17} \pm 0.02$, consistent with no asymmetry.
\end{abstract}

\pacs{13.25.Ft, 13.25.Hw, 14.40.Nd}

\maketitle

%%%% >>>> keep the final version single-spaced
\tighten

{\renewcommand{\thefootnote}{\fnsymbol{footnote}}}
\setcounter{footnote}{0}

%%%
%%% Instroduction
%%%
Evidence of charmless $B \to \phi \phi K$ decays has been reported by the Belle collaboration using $85 \times 10^6$ $B\overline{B}$ pairs~\cite{Huang}.
In the Standard Model (SM), this decay channel requires the creation of an additional final $s \overline{s}$ quark pair in a $b \to s \overline{s} s$ process, such as $B \to \phi K$.
Therefore, the study of $B \to \phi \phi K$ provides useful information for 
understanding quark fragmentation in $B$ decays.  Moreover, our previous  
study also reported results for the decays $B \to J/\psi K (\eta_c K)$ with
the $J/\psi (\eta_c)$ decaying into four kaons in the final state, which can
proceed with $\phi$ mesons in the intermediate state.       
It has been suggested that large direct $CP$ violation up to 40\% 
is possible in $B\to\eta_c K\to\phi\phi K$ if there is a sizable $b\to s$
transition with a non-SM $CP$-violating phase that interferes with the
decay amplitude via the $\eta_c$ resonance ~\cite{Hazumi}. 
 
Recently the BaBar collaboration has reported  results of a study of
$B\to \phi\phi K$~\cite{BaBar}. The  branching fraction for 
$B^\pm\to \phi\phi K^\pm$ that they obtained  is around three times 
larger than our previous 
measurement~\cite{Huang}. Here we present  improved measurements of 
$B \to \phi \phi K$ decays with not only larger 
statistics but also proper consideration of the non-resonant $K^+K^-$ 
contribution
underneath the $\phi$ resonance. 
The analysis is based on a data sample of $414~{\rm fb}^{-1}$ containing
 449 $\times 10^6 B\overline{B}$ pairs. The data were collected with the Belle 
detector at the KEKB asymmetric-energy $e^+e^-$ (3.5 on 8~GeV) collider~\cite{KEKB} operating at the $\Upsilon(4S)$ resonance.\par

The Belle detector is a large-solid-angle magnetic spectrometer that consists of a silicon vertex detector (SVD), a 50-layer central drift chamber (CDC), 
an array of aerogel threshold Cherenkov counters (ACC), a barrel-like arrangement of time-of-flight scintillation counters (TOF), and an electromagnetic calorimeter comprised of CsI(Tl) crystals (ECL) located inside a superconducting solenoid coil that provides a 1.5~T magnetic field.
An iron flux-return located outside  the coil is instrumented to detect $K_L^0$ mesons and to identify muons (KLM).
The detector is described in detail elsewhere~\cite{Belle}.
Two inner detector configurations were used.
A 2.0 cm radius beampipe and a 3-layer silicon vertex detector (SVD1) were used for the first sample of 152 $\times 10^6 B\overline{B}$ pairs, while a 1.5 cm radius beampipe, a 4-layer silicon detector (SVD2) and a small-cell inner drift chamber were used to record the remaining 297 $\times 10^6 B\overline{B}$ pairs~\cite{Ushiroda}.\par

%%%
%%% Event selection and B reconstruction
%%%
Charged kaons are required to have impact parameters within $\pm$2 cm of the interaction point (IP) along the $z$-axis (antiparallel to the positron direction) and within 0.2 cm in the transverse plane.
Each track is identified as a kaon or a pion according to a $K/\pi$ likelihood ratio, $\mathcal{R}(K/\pi) = \mathcal{L}_K/(\mathcal{L}_K+\mathcal{L}_\pi)$, where $\mathcal{L}_K/\mathcal{L}_\pi$ is the likelihood of kaons/pions derived from the responses of TOF and ACC systems and the energy loss measurements from the CDC.
The likelihood ratio is required to exceed 0.6 for kaon candidates; within the momentum range of interest, this requirement is 88\% efficient for kaons and has a misidentification rate for pions of 8.5\%.
Neutral kaons are reconstructed via the decay $K^{0}_{S} \to \pi^{+} \pi^{-}$ and have an invariant mass 0.482 GeV/$c^2 < M_{\pi^{+} \pi^{-}} <$ 0.514 GeV/$c^2$ ($\pm4 \sigma$ mass resolution).
The $\pi^{+} \pi^{-}$ vertex is required to be displaced from the IP and the flight direction must be consistent with a $K^{0}_{S}$ that originated from the IP.
The required displacement increases with the momentum of the $K^0_S$ candidate.\par

$B$ meson candidates are reconstructed in the five-kaon final state.
Two kinematic variables are used to distinguish signal candidates from backgrounds: the beam-energy constrained mass $M_{\rm bc} = \sqrt{E^{2}_{\rm beam} - |\vec{P}_{\rm recon}|^{2}}$ and the energy difference $\Delta E = E_{\rm recon} - E_{\rm beam}$, where $E_{\rm beam}$ is the beam energy, and $E_{\rm recon}$ and $\vec{P}_{\rm recon}$ are the reconstructed energy and momentum of the signal candidate in the $\Upsilon(4S)$ rest frame. The resolution of $M_{\rm bc}$ is approximately 2.8 MeV/$c^2$, dominated by the beam energy spread, while the $\Delta E$ resolution is around 10 MeV.
Candidates with five kaons within the region $|\Delta E| <$ 0.2 GeV and 5.2 GeV/$c^2 < M_{\rm bc}$ are selected for further consideration.
The signal region is defined as 5.27 GeV/$c^2 < M_{\rm bc} <$ 5.29 GeV/$c^2$ and $|\Delta E| <$ 0.05 GeV.\par

%%%
%%% Background suppression
%%%
The dominant backgrounds are $e^{+}e^{-} \to q\overline{q}$ ($q=u, d, s, c$) continuum events.
Event topology and $B$ flavor tagging are used to distinguish the jet-like continuum events and the spherically distributed $B\overline{B}$ events. 
Seven event-shape variables are combined into a single Fisher discriminant~\cite{fisher}.
The Fisher variables include the angle between the thrust axis of the $B$ candidate and the thrust axis of the rest of the event ($\cos\theta_{T}$),  five modified Fox-Wolfram moments~\cite{SFW}, and a measure of the momentum transverse to the event thrust axis ($S_\perp$)~\cite{spher}.
Two other variables that are uncorrelated with the Fisher discriminant and help to distinguish signal from the continuum are $\cos\theta_{B}$, where $\theta_{B}$ is the angle between the $B$ flight direction and the beam direction in the $e^+e^-$ center-of-mass frame, and $\Delta z$, the $z$ vertex difference between the signal $B$ candidate and its accompanying $B$.
We form signal and background probability density functions (PDFs) for the Fisher discriminant, $\cos\theta_{B}$ and $\Delta z$ using the signal Monte Carlo (MC) events and sideband data (5.2 GeV/$c^2 < M_{\rm bc} <$ 5.26 GeV/$c^2$), respectively. 
The products of the PDFs for these variables give signal and background likelihoods $\mathcal{L}_{S}$ and $\mathcal{L}_{BG}$ for each candidate, allowing a selection to be applied to the likelihood ratio $\mathcal{R}=\mathcal{L}_{S}/(\mathcal{L}_{S}+\mathcal{L}_{BG})$.\par

Additional background discrimination is provided by the quality of the $B$ flavor tagging of the  accompanying $B$ meson.
The standard Belle flavor tagging package~\cite{TaggingNIM} gives two outputs: a discrete variable indicating the flavor of the tagging $B$ and dilution factor $r$, which ranges from zero for no flavor information to unity for unambiguous flavor assignment.
The continuum background is reduced by applying a selection requirement on the
variable  $\mathcal{R}$ for events in each $r$ region according to the figure of merit defined as $N_{S}/\sqrt{N_{S} + N_{BG}}$, where $N_{S}$ denotes the expected $\phi \phi K$ signal yield based on MC simulation and the branching fraction reported in our previous measurements, and $N_{BG}$ denotes the expected $q\overline{q}$ yields from sideband data.
This requirement removes (61-81)\% of the continuum background while retaining (80-92)\% of the signal, and depends on the decay channel ($\phi \phi K$ or $\phi \phi K^0$) and the SVD configuration during the measurement (SVD1 or SVD2).
Backgrounds from other $B$ decays are investigated using a large MC sample and are found to be negligible after the $\mathcal{R}$ requirement.\par

%%%
%%% Signal extraction
%%%
The signal yields are extracted by applying an unbinned extended maximum likelihood (ML) fit to the events with $M_{\rm bc} >$ 5.2 GeV/$c^2$ and 
$|\Delta E| <$ 0.2 GeV.
For the $\phi\phi K^\pm$ mode, we simultaneously obtain the yield and the partial rate asymmetry $\mathcal{A}_{CP}$ using the likelihood, defined as:  

\begin{eqnarray}
%\mathcal{L} = {\rm exp}[-(N_S + N_{BG})] \prod_i^N (\sum_j \frac{1}{2}[1-q_i\cdot {\mathcal A}^j_{CP}] N_j P_i^j(M_{\rm bc},\Delta E)),
\mathcal{L} = e^{-(N_S + N_{BG})} \prod_i^N (\sum_j \frac{1}{2}[1-q_i\cdot {\mathcal A}^j_{CP}] N_j P_i^j),
\label{eq: ML} 
\end{eqnarray}
where $i$ is the identifier of the $i$-th event, $j$ indicates signal ($S$) 
or background ($BG$), $P$ is the two-dimensional PDF of $M_{\rm bc}$ and $\Delta E$, and $q$ indicates the $B$ meson flavor, $+1$ for $B^+$ and $-1$ for $B^-$, respectively.
For neutral $B$ events, the factor $\frac{1}{2}[1-q_i\cdot {\mathcal A}_{CP}]$ in Eq. (\ref{eq: ML}) is replaced by 1.
%%%
%%% version 1.6 Tom modified English
%%%
The $M_{\rm bc}$ PDFs are modeled by a Gaussian function for signals and an ARGUS function~\cite{argus} for the continuum, while a Gaussian is used to describe the signal $\Delta E$ and a second-order Chebyshev polynomial is used for the background $\Delta E$ distribution. 
The parameters of the PDFs are determined using high-statistics MC samples and sideband data for signal and background shapes, respectively.
%%%
%%% version 1.6 Tom modified English
%%%
The signal PDFs are calibrated by comparing the $M_{\rm bc}$ and $\Delta E$ distributions of the $B^{+} \to \overline{D}{}^{0}(K^+\pi^-\pi^-\pi^+) \pi^{+}$ data with the MC expectation.\par

\begin{figure}[htb]
\includegraphics[scale=0.45]{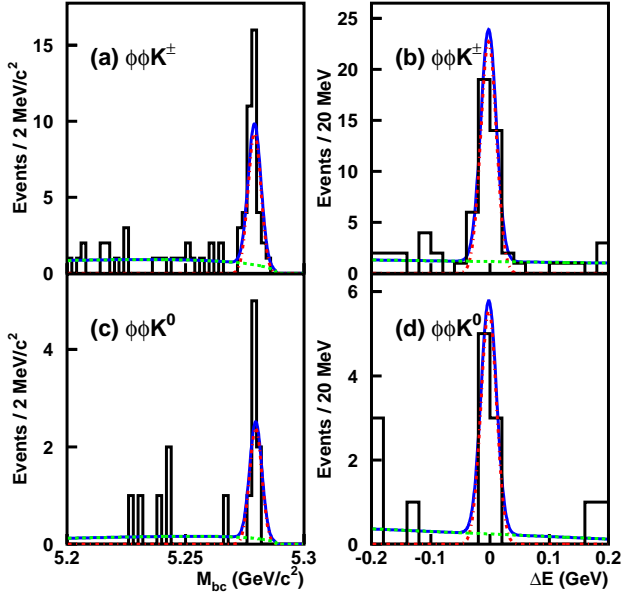}
\caption{Distributions of $M_{\rm bc}$ and $\Delta E$ for the decay modes $B^{\pm} \to \phi \phi K^{\pm}$ (a,b) and $B^{0} \to \phi \phi K^{0}$ (c,d), with $\phi \phi$ invariant mass less then 2.85 GeV/$c^2$.
The open histograms are the data, the solid curves show the result of the fit,  the dash-dotted curves represent the signal contributions and the dashed curves show the continuum background contributions.}
\label{fig: phiphik_fig1}
\end{figure}

\begin{figure}[htb]
\includegraphics[scale=0.45]{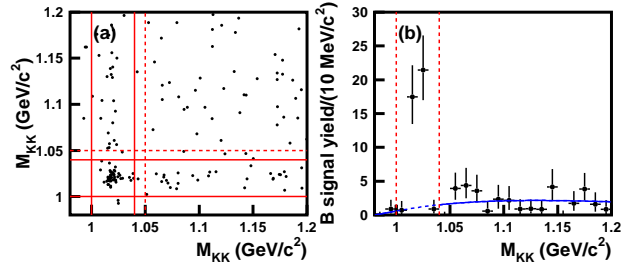}
\caption{(a) The distribution of $M^1_{K^+K^-}$ vs. $M^2_{K^+K^-}$ for the $K^+K^-K^+K^-K^\pm$ candidates in the $M_{\rm bc}-\Delta E$ signal box with $M_{K^+K^-} <$ 1.2 GeV/$c^2$.
The two $K^+K^-$ bands indicate the $\phi$ mass region (1.0 GeV/$c^2 < M_{K^+K^-} <$ 1.04 GeV/$c^2$).
The horizontal and vertical dashed lines are located at $M_{K^+K^-}$ = 1.05 GeV/$c^2$.
The rectangle on the upper right is the $\phi \phi$ sideband region; events in this region are used to estimate the non-resonant $B\to 5K$ contribution.
(b) $B$ signal yield as a function of the $M_{K^+K^-}$ after requiring the other $K^+K^-$ pair to have a mass in the $\phi$ mass region.
%%%
%%% version 1.6 Tom modified English
%%%
A threshold function is used to fit the data; events with 1.0 GeV/$c^2 < M_{K^+K^-} <$ 1.04 GeV/$c^2$ are excluded when the fit is performed.}
\label{fig: phiphik_fig2}
\end{figure}

We search for charmless $B\to \phi\phi K$ decays by requiring the $\phi \phi$ invariant mass ($M_{\phi \phi}$) to be less than 2.85  GeV/$c^2$, the region below  charmonium threshold.
Candidate $\phi$ mesons are identified by requiring the invariant masses of $K^+K^-$ pairs ($M_{K^+K^-}$) to be in the range 1.0 GeV/$c^2$ to 1.04 GeV/$c^2$ ($\pm 4.6\sigma$).
Figure \ref{fig: phiphik_fig1} shows the $M_{\rm bc}$ and $\Delta E$ projections with the fit curves superimposed. 
Clear signals appear in both $B^\pm$ and $B^0$ modes with signal yields of $37.0^{+6.7}_{-6.0}$ and $7.8^{+3.2}_{-2.5}$, respectively.
Although $K^+K^-$ candidates are required to lie in the $\phi$ mass region, non-$\phi$ backgrounds may also contribute.
Figure \ref{fig: phiphik_fig2}(a) shows the $M^1_{K^+K^-}$ vs. $M^2_{K^+K^-}$ distributions for $(K^+K^-K^+K^-)K^\pm$ candidates in the signal region, where the two $K^+K^-$ pairs are required to have invariant masses less than 1.2 GeV/$c^2$.
Events in the two $\phi$ bands are used to estimate the $B^\pm\to\phi K^+K^-K^\pm$ contribution.
%%%
%%% version 1.6 Tom modified English
%%%
Figure \ref{fig: phiphik_fig2}(b) shows the $B$ signal yields~\cite{B signal yields} as a function of  the $K^+K^-$ invariant mass after requiring the other $K^+ K^-$ pair to have a mass in the $\phi$ mass region.
The $B$ signal yields are fitted with a threshold function in the region 0.98 GeV/$c^2 < M_{K^+K^-} <$ 1.20 GeV/$c^2$, excluding the $\phi$ mass region (1.00 GeV/$c^2 < M_{K^+K^-} <$ 1.04 GeV/$c^2$).
%%%
%%% version 1.6 Tom modified English
%%%
The size of the  non-$\phi$ contribution is estimated by interpolating the $B$ yields in the $\phi$ sideband region to the $\phi$ mass region.
%%%
%%%
Since events in the two $\phi$ bands contain both true $\phi$ mesons and non-resonant $K^+K^-$ pairs, the area underneath the $\phi$ mass region in Fig. \ref{fig: phiphik_fig2}(b) also includes  the $\phi K^+K^-K^\pm$ contribution and, due to combinatorics, a double counted contribution from the non-resonant 
$5 K$ component.
%%%
%%%
The contribution of $B \to 5K$ is estimated by extrapolating the $B$ signal yield in the upper right corner of the dashed region in Fig. \ref{fig: phiphik_fig2}(a) to the $\phi$ mass region.
The fraction of non-$\phi \phi K$ events in the $\phi$ mass region as obtained from both contributions is thus ($7\pm 4$)\%.
The same procedure is applied to the $\phi\phi K^0$ sample; here we obtain a fraction of ($7 \pm 9$)\%.\par

Table \ref{table: phiphik results} summarizes the $\phi \phi K$ results after subtracting the non-$\phi \phi K$ contribution.
Signal efficiencies are obtained by generating $\phi \phi K$ MC events, where the same $M_{\phi \phi} <$ 2.85 GeV/$c^2$ requirement is applied.
Systematic uncertainties in the fit are obtained by performing fits in which the signal peak positions and resolutions of the signal PDFs are successively varied by $\pm 1 \sigma$.
The quadratic sum of each deviation from the central value of the fit gives the systematic uncertainty of the fit.
For each systematic check, the statistical significance is taken as $\sqrt{-2\ln(\mathcal{L}_{\rm feeddown}/\mathcal{L}_{\rm max})}$, where $\mathcal{L}_{\rm feeddown}$ and $\mathcal{L}_{\rm max}$ are the likelihoods at the expected non-$\phi\phi K$ yields and the best fit, respectively. The change in significance 
that arises from uncertainties in the signal PDFs is negligible ($<1\%$). 
We choose the 
significance
calculated after increasing the non-$\phi\phi K$ yield by its $ 1 \sigma$ 
statistical uncertainty as our significance including  systematic 
uncertainty.
The numbers of $B^+B^-$ and $B^0\overline{B}{}^0$ pairs are assumed to be equal.\par

%%%
%%% Results table
%%%
%\begin{table}[htb]
\begin{table*}[htb]
%%%
%%%
%%%
\caption{Mode, yield, efficiency including secondary branching fractions, branching fraction for $B \to \phi \phi K$ and related charmonium decays.}
\label{table: phiphik results}
\begin{tabular}
{@{\hspace{3mm}}l@{\hspace{3mm}}@{\hspace{3mm}}c@{\hspace{3mm}}@{\hspace{3mm}}c@{\hspace{3mm}}@{\hspace{3mm}}c@{\hspace{3mm}}@{\hspace{3mm}}c@{\hspace{3mm}}@{\hspace{3mm}}c@{\hspace{3mm}}}
\hline \hline
Mode & Yield & Efficiency(\%) & $\mathcal{B} (10^{-6})$ \\
\hline
$B^{\pm} \to \phi \phi K^{\pm}$  ($M_{\phi \phi} <$ 2.85 GeV/$c^2$) & $34.2^{+6.4}_{-5.8}$ & 2.41 & $3.2^{+0.6}_{-0.5} \pm 0.3$ \\
$B^{0} \to \phi \phi K^{0}$ ($M_{\phi \phi} <$ 2.85 GeV/$c^2$) & $7.3^{+3.0}_{-2.4}$ & 0.69 & $2.3^{+1.0}_{-0.7} \pm 0.2$ \\
$B^\pm \to \eta_c K^\pm$, $\eta_c \to \phi \phi$ & $27.9^{+7.3}_{-6.9}$ & 2.72 & $2.3\pm0.6\pm 0.2$ \\
$B^\pm \to \eta_c K^\pm$, $\eta_c \to \phi K^+K^-$ & $60.3^{+12.2}_{-11.8}$ & 4.85 & $2.8^{+0.6}_{-0.5}\pm 0.2$ \\
$B^\pm \to \eta_c K^\pm$, $\eta_c \to 2(K^+K^-)$ & $105.7^{+26.1}_{-20.7}$ & 9.93 & $2.4^{+0.6}_{-0.5}\pm 0.2$ \\
$B^\pm \to J/\psi K^\pm$, $J/\psi \to \phi K^+K^-$ & $26.3^{+6.9}_{-6.1}$ & 4.67 & $1.3\pm 0.3\pm 0.2$ \\
$B^\pm \to J/\psi K^\pm$, $J/\psi \to 2(K^+K^-)$ & $36.0^{+7.7}_{-7.3}$ & 9.41 & $0.85^{+0.18}_{-0.17}\pm 0.10$ \\ 
\hline \hline
\end{tabular}
\end{table*}

%%%
%%%  Systematics
%%%
The systematic uncertainty resulting from the $\mathcal{R}$ requirement is studied by checking the data-MC efficiency ratio using the $B^{+} \to \overline{D}{}^{0}(\to K^{+} \pi^{-} \pi^{-} \pi ^{+})\pi^{+}$ sample.
%%%
%%% version 1.5
%%%
The corresponding systematic error is 2.7-2.8\% and again depends on the decay channel and SVD geometry.
The systematic errors on the charged track reconstruction are estimated to be around $1$\% per track using  partially reconstructed $D^*$ events.
Therefore, the tracking systematic error is 5\% (five tracks) for the $\phi\phi K^\pm$ mode and 4\% for the $\phi\phi K^0$ mode (excluding $K^0_S$ reconstruction).
The kaon identification efficiency is studied using samples of inclusive $D^{*+}\to D^0\pi^+, D^0\to K^-\pi^+$ decays.          
The $K_S^0$ reconstruction is verified by comparing the  ratio of $D^+\to K_S^0\pi^+$ and $D^+\to K^-\pi^+\pi^+$ yields.
The resulting $K_S^0$ detection systematic error is 4.9\%.
The uncertainty in the number of $B\overline{B}$ events is 1.4\%. 
The final systematic error is obtained  by summing  all correlated errors linearly and then quadratically summing the uncorrelated errors.\par

%%%
%%% Summary of charmless phiphik
%%%
After subtracting the non-$\phi\phi K$ contribution, the branching fractions for charmless $B \to \phi \phi K$ decays are $\mathcal{B}(B^{\pm} \to \phi \phi K^{\pm}) = (3.2^{+0.6}_{-0.5} \pm 0.3) \times 10^{-6}$ with a 9.5$\sigma$ significance and $\mathcal{B}(B^{0} \to \phi \phi K^{0}) = (2.3^{+1.0}_{-0.7} \pm 0.2) \times 10^{-6}$ with a 4.7$\sigma$ significance.
The measured charge asymmetry for $B^{\pm} \to \phi \phi K^{\pm}$ decay is $0.01^{+0.19}_{-0.16} \pm 0.02$.
The first error is statistical and the second is systematic.\par

\begin{figure}[htb]
%%%
\includegraphics[scale=0.45]{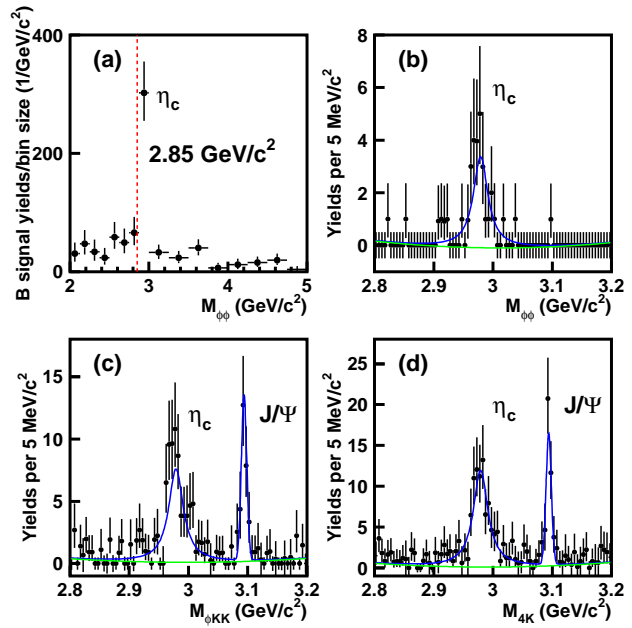}
\caption{$B^\pm$ signal yield as a function of (a,b) $M_{\phi \phi}$, (c) $M_{\phi K^{+}K^{-}}$ and (d) $M_{4K}$.
In (a) we use different bin sizes for $M_{\phi \phi}$ less than 3 GeV/$c^2$ and greater than 3 GeV/$c^2$.
The subset with $M_{\phi \phi}$ from 2.8 GeV/$c^2$ to 3.2 GeV/$c^2$ is shown in (b). The $J/\psi$ signal PDF is modeled with a Gaussian function while the
 $\eta_c$ PDF is described by a Breit-Wigner function convoluted with a Gaussian resolution function, which has the same Gaussian width as the $J/\psi$ PDF. 
The solid curves in (b,c,d) show the results of the fit and the contributions 
(second  order polynomial) not from the $J/\psi$ and $\eta_c$.}
\label{fig: phiphik_fig3}
\end{figure}

%%%
%%% The related charmonium decays
%%%
It is of interest to search for possible $\phi \phi$ resonances above  
charmonium  threshold.
Figure \ref{fig: phiphik_fig3}(a) shows the $B$ signal yield divided by the bin size as a function of $M_{\phi \phi}$ if the $M_{\phi \phi} <$ 2.85 GeV/$c^2$ requirement is not applied.
%%%
%%% version 1.6 Tom modified English
%%%
There is no enhancement in the high $\phi \phi$ mass region except for the $\eta_{c}$ peak near 3 GeV/$c^2$.
Reference \cite{Hazumi} suggests the  possibility of a large $CP$ asymmetry, which could arise from the interference between $B^\pm \to \phi \phi K^\pm$ and $B^\pm \to \eta_c(\to \phi \phi) K^\pm$ decays.
Events with $\phi \phi$ invariant mass within $\pm 40$ MeV/$c^2$ of the nominal $\eta_c$ mass are selected to investigate this asymmetry.
The measured $CP$ asymmetry is $0.15^{+0.16}_{-0.17} \pm 0.02$, which is consistent with no asymmetry.\par

%%%
%%% version 1.6 Tom modified English
%%%
We study possible charmonium states by measuring the $B$ yield with $M_{4K}$ between 2.8 GeV/$c^2$ and 3.2 GeV/$c^2$.
Since $\eta_c$ and $J/\psi$ mesons may decay to $\phi K^+K^-$ and  $2(K^+K^-)$, mass fits are performed with and without the requirement that one or both
 $K^+K^-$ pairs lie in the $\phi$ mass region.
As shown in Fig. \ref{fig: phiphik_fig3}, clear $\eta_c$ and $J/\psi$ resonances are visible in the $\phi K^+K^-$ and $4K$ samples while only an $\eta_c$ peak appears in the $\phi \phi$ mode.\par

We obtain the signal yields for $B^\pm\to \eta_c K^\pm$ and $B^\pm\to J/\psi K^\pm$ by performing $\chi^2$ fits with  asymmetric errors to the $M_{\phi\phi}$, $M_{\phi K^+K^-}$ and $M_{4K}$ invariant mass distributions, which are presented in Figs. \ref{fig: phiphik_fig3}(b, c, d).
%%%
%%%
The $J/\psi$ signal PDF is modeled with a Gaussian function while the $\eta_c$ PDF is described by a Breit-Wigner function convoluted with a Gaussian resolution function, which has the same Gaussian width as the $J/\psi$ PDF.
Since sizable signals are observed in the $4K$ mode, the parameters are determined using the $4K$ sample and the same signal PDFs are then applied to the $\phi K^+K^-$ and $\phi \phi$ samples.
%%%
%%%
The obtained Gaussian width is measured to be $4.5^{+1.9}_{-1.3}$ MeV/$c^2$.
The resulting signal yields are summarized in Table \ref{table: phiphik results}.
The peak positions obtained for the $\eta_c$ and $J/\psi$ are 
$2.979\pm 0.002\pm 0.001$ GeV/$c^2$ and $3.094\pm 0.001$ GeV/$c^2$, respectively, consistent with the nominal $\eta_c$ and $J/\psi$ masses.
The $\eta_c$ Breit-Wigner width is measured to be $29.8^{+12.2}_{-8.5} \pm 0.1$ MeV/$c^2$, where the central value is consistent with the world average \cite{PDG} and the second error is due to the uncertainty of the Gaussian width for
the mass resolution.
The validity of determining $B$ signal yields from a constrained $\chi^2$ fit with an asymmetric error is verified by toy MC.\par

For the $\phi K^+K^-$ and $\phi \phi$ modes, the non-$\phi$ contribution is determined from the $B$ signal yields for events with one $K^+K^-$ pair in the $\phi$ sideband region (1.05 GeV/$c^2 <M_{K^+K^-} <$ 1.09 GeV/$c^2$) and the $4K$ and $\phi K^+K^-$ masses  in the charmonium resonance region, respectively.
%%%
%%% Version 1.0
%%%
We find $3.0^{+2.1}_{-1.4}$ events in the $\eta_c \to \phi \phi$ mode, $6.4^{+5.4}_{-4.5}$ events in the $\eta_c \to \phi K^+ K^-$ mode, and $3.5^{+3.6}_{-2.6}$ events in the $J/\psi \to \phi K^+ K^-$ mode.
%After subtracting the feed-down yields, we obtain the results listed in Table \ref{table: phiphik results}.\par
%%%
%%%
For the yields of these modes, listed in Table \ref{table: phiphik results}, the corresponding feed-down yields have been subtracted.\par

Signal efficiencies are determined using signal MC and their  systematic 
uncertainties are similar to what was described in the charmless 
$\phi\phi K$ part.
%%%
Systematic uncertainties in the fitting are obtained by performing fits in which the signal peak positions, the resolutions of the signal PDF's and the width of the Gaussian resolution function convoluted with the $\eta_c$ PDF are successively varied by $\pm \sigma$.
%%%
%%%
The quadratic sum of all deviations gives the systematic error of the fit.
The  products of the branching fractions for various decays are listed in
Table \ref{table: phiphik results}.
Since the probabilities of  $\eta_c$ and $J/\psi$ decays to $4K, \phi KK$ and $\phi\phi$ final states are measured with poor accuracy, 
we can use the values from Table \ref{table: phiphik results} as well as 
the world average  branching fractions 
$\mathcal{B}(B^{\pm} \to \eta_{c} K^{\pm}) = (9.1 \pm 1.3) \times 10^{-4}$ and $\mathcal{B}(B^{\pm} \to J/\psi K^{\pm}) = (1.007 \pm 0.035) \times 10^{-3}$ \cite{PDG} to determine independently the corresponding
 branching fractions for the $\eta_c$ and $J/\psi$; the results are shown in 
Table \ref{table: subdecay BF}. The world average values given  above are 
based on datasets, which also include Belle measurements \cite{belle_ch}; 
however, our estimate shows that effects of possible correlations are 
negligible.
We also provide the ratios of branching fractions as shown in 
Table~\ref{tab:ratio}. The systematic uncertainties are predominately due to 
 the Gaussian width for the mass resolution; other systematic uncertainties 
 either cancel out in the ratios or too small to be considered. 
 
\begin{table}[htb]
\caption{The measured branching fractions of secondary charmonium decays.}
\label{table: subdecay BF}
\begin{tabular}
{@{\hspace{0.5cm}}l@{\hspace{0.5cm}}@{\hspace{0.5cm}}c@{\hspace{0.5cm}}@{\hspace{0.5cm}}c@{\hspace{0.5cm}}}
\hline \hline
Decay mode & $\mathcal{B}$ \\
\hline
$\eta_{c} \to \phi \phi$ & $(2.5^{+0.7}_{-0.6} \pm 0.4) \times 10^{-3}$ \\
$\eta_{c} \to \phi K^{+}K^{-}$ & $(3.0 \pm 0.6 \pm 0.5) \times 10^{-3}$ \\
$\eta_{c} \to 2(K^{+}K^{-})$ & $(2.6^{+0.6}_{-0.5} \pm 0.4) \times 10^{-3}$ \\
$J/\psi \to \phi K^{+}K^{-}$ & $(1.2 \pm 0.3 \pm 0.2) \times 10^{-3}$ \\
$J/\psi \to 2(K^{+}K^{-})$ & $(8.5^{+1.8}_{-1.7} \pm 1.0) \times 10^{-4}$ \\
\hline \hline
\end{tabular}
\end{table}

\begin{table}[htb]
\caption{Ratios of branching fractions for the $\eta_c$ and $J/\psi$ decays.}
\label{tab:ratio}
\begin{tabular}
{@{\hspace{0.5cm}}c@{\hspace{0.5cm}}@{\hspace{0.5cm}}c@{\hspace{0.5cm}}@{\hspace{0.5cm}}c@{\hspace{0.5cm}}}
\hline \hline
Mode & Ratio \\ \hline
$\frac{{\cal B}(\eta_c \to \phi\phi)}{{\cal B}(\eta_c\to 2(K^+K^-))}$ &
 $0.96^{+0.20}_{-0.24}\pm 0.02$ \\
$\frac{{\cal B}(\eta_c \to \phi K^+ K^-)}{{\cal B}(\eta_c\to 2(K^+K^-))}$ &
$1.17^{+0.33}_{-0.36}\pm 0.01$ \\
$\frac{{\cal B}(J/\psi\to \phi K^+ K^-)}{{\cal B}(J/\psi\to 2(K^+K^-))}$ &
$1.47^{+0.49+0.06}_{-0.46-0.08}$ \\ \hline \hline 
\end{tabular}
\end{table}

%%%
%%% Summary
%%%
In summary, we have observed the charmless decay $B^{\pm} \to \phi \phi K^{\pm}$ and  evidence for $B^{0} \to \phi \phi K^{0}$.
We also report the $CP$ asymmetry of the charged decay and measurements of other closely related charmonium decays.
The results are consistent with our previous measurements~\cite{Huang} 
and supersede them, and have considerably better precision due to the increase in statistics. The obtained
branching fraction of charmless $B^\pm\to \phi \phi K^\pm$ decay is $3.1\sigma$ 
smaller than the measurement by the BaBar collaboration~\cite{BaBar}.
For the charmonium decays, we find that the decay $\eta_c\to \phi\phi$ 
contributes around 42\% and 24\% of the $\eta_c\to \phi K^+K^-$ and $\eta_c \to
2(K^+K^-)$ events, respectively. The latter is consistent with an early 
Belle measurement~\cite{gg}  using $\eta_c$ events produced in two 
photon collisions.    
For both the $\eta_c$ and $J/\psi$ decays into four charged kaons, the 
$\phi K^+K^-$ contribution dominates.

%----------- Long version, for most papers ----------- 
%We thank the KEKB group for the excellent operation of the
%accelerator, the KEK cryogenics group for the efficient
%operation of the solenoid, and the KEK computer group and
%the National Institute of Informatics for valuable computing
%and Super-SINET network support. We acknowledge support from
%the Ministry of Education, Culture, Sports, Science, and
%Technology of Japan and the Japan Society for the Promotion
%of Science; the Australian Research Council and the
%Australian Department of Education, Science and Training;
%the National Science Foundation of China and the Knowledge 
%Innovation Program of the Chinese Academy of Sciences under 
%contract No.~10575109 and IHEP-U-503; the Department of Science 
%and Technology of India; the BK21 program of the Ministry of Education of
%Korea, and the CHEP SRC program and Basic Research program 
%(grant No. R01-2005-000-10089-0) of the Korea Science and
%Engineering Foundation; the Polish State Committee for
%Scientific Research under contract No.~2P03B 01324; the
%Ministry of Science and Technology of the Russian
%Federation; the Slovenian Research Agency;  
%the Swiss National Science Foundation; the National Science Council and
%the Ministry of Education of Taiwan; and the U.S.\
%Department of Energy.

%-------- Short version, if necessary, for PRL -----------
%(A short version is also available on 
%    ``http://belle.kek.jp/\~\ kinosh/private/pub/ack.txt'',
%but please use the long version for conference papers. ) 
We thank the KEKB group for excellent operation of the
accelerator, the KEK cryogenics group for efficient solenoid
operations, and the KEK computer group and
the NII for valuable computing and Super-SINET network
support.  We acknowledge support from MEXT and JSPS (Japan);
ARC and DEST (Australia); NSFC and KIP of CAS (China); 
DST (India); MOEHRD, KOSEF and KRF (Korea); 
KBN (Poland); MES and RFAAE (Russia); ARRS (Slovenia); SNSF (Switzerland); 
NSC and MOE (Taiwan); and DOE (USA).


\begin{thebibliography}{99}
\bibitem{Huang} 
H.C.~Huang {\it et al.} (Belle Collaboration),
Phys. Rev. Lett. {\bf 91}, 241802 (2003).

\bibitem{Hazumi}
M.~Hazumi,
Phys. Lett. B {\bf 583}, 285 (2004).

\bibitem{BaBar}
B. Aubert {\it et al.} ({\it BABAR} Collaboration),
Phys. Rev. Lett. {\bf 97}, 261803 (2006).

\bibitem{KEKB}
S.~Kurokawa and E.~Kikutani, Nucl. Instr. and. Meth. A {\bf 499}, 1 (2003),
and other papers included in this volume.

\bibitem{Belle}
A.~Abashian {\it et al.} (Belle Collaboration),
Nucl. Instr. and Meth. A {\bf 479}, 117 (2002).

\bibitem{Ushiroda} Y. Ushiroda,
Nucl. Instr. and Meth. A {\bf 511} 6 (2003); Z.~Natakaniec {\it et al.},
(Belle SVD2 Group), Nucl. Instr. and Meth. A {\bf 560}, 1 (2006).


\bibitem{fisher}
R. A. Fisher, Ann. Eugenics {\bf 7}, 179 (1936).


\bibitem{SFW}
The Fox-Wolfram moments were introduced in G.~C.~Fox and S.~Wolfram, Phys. Rev. Lett. {\bf 41}, 1581 (1978).
The Fisher discriminant used by Belle, based on modified Fox-Wolfram moments (SFW), is described in K.~Abe {\it et al.} (Belle Collaboration), Phys. Rev. Lett. {\bf 87}, 101801 (2001) and K.~Abe {\it et al.} (Belle Collabboration.), Phys. Lett. B {\bf 511}, 151 (2001).
%The Fisher discriminant used by Belle, based on modified Fox-Wolfram moments (SFW), is described in K.~Abe {\it et al.} (Belle Collab.), Phys. Rev. Lett. {\bf 87}, 101801 (2001) and K.~Abe {\it et al.} (Belle Collab.), Phys. Lett. {\bf B 511}, 151 (2001).

\bibitem{spher} R. Ammar {\it et al.} (CLEO Collaboration),
 Phys. Rev. Lett. {\bf 71}, 674 (1993).

\bibitem{TaggingNIM}
H. Kakuno {\it et al.}, Nucl. Instr. and Meth. A {\bf 533}, 516 (2004).

\bibitem{argus}
H. Albrecht {\it et al.} (ARGUS Collaboration), Phys. Lett. B {\bf 229}, 304 (1989).

\bibitem{B signal yields}
Hereafter in this paper, the $B$ signal yield is obtained from 2D $M_{\rm bc}-\Delta E$ fits to the events in each bin of the plot.

\bibitem{PDG}
W.-M. Yao {\it et al.} (Particle Data Group), 
Journal of Physics G {\bf 33}, 1 (2006) and 2007 partial update for
 edition 2008.

\bibitem{belle_ch}
F. Fang {\it et al.} (Belle Collaboration), Phys. Rev. Lett. {\bf 90}, 
071801 (2003); K. Abe {\it et al.} (Belle Collaboration), Phys. Rev. D 
{\bf 67}, 032003 (2003). 

\bibitem{gg}
S. Uehara {\it et al.} (Belle Collaboration), Euro. Phys. Jour. C {\bf 53}, 1 (2008).


\end{thebibliography}
\end{document}